\documentclass[journal]{IEEEtran}
\usepackage{blindtext}
\usepackage{graphicx}

\ifCLASSINFOpdf
\else
\fi
\usepackage[cmex10]{amsmath}
\begin{document}
%
\title{Observation of Spatio-temporal Instability of Femtosecond Pulses in Normal Dispersion Multimode Graded-Index Fiber}
%
%
%

\author{U\u{g}ur Te\u{g}in
        and B\"{u}lend Orta\c{c}
\thanks{U. Te\u{g}in and B. Orta\c{c} are with National Nanotechnology Research Center (UNAM) and Institute of Materials Science and Nanotechnology, Bilkent University, 06800 Bilkent, Ankara, Turkey (e-mails: ugur.tegin@bilkent.edu.tr and ortac@unam.bilkent.edu.tr)}
\thanks{}}

\markboth{}%
{Shell \MakeLowercase{\textit{et al.}}: Bare Demo of IEEEtran.cls for Journals}
%



\maketitle

\begin{abstract}
We study the spatio-temporal instability generated by a universal unstable attractor in normal dispersion graded-index multimode fiber (GRIN MMF) for femtosecond pulses. Our results present the generation of geometric parametric instability (GPI) sidebands with ultrashort input pulse for the first time. Observed GPI sidebands are 91 THz detuned from the pump wavelength, 800 nm. Detailed analysis carried out numerically by employing coupled-mode pulse propagation model including optical shock and Raman nonlinearity terms. A simplified theoretical model and numerically calculated spectra are well-aligned with experimental results. For input pulses of 200-fs duration, formation and evolution of GPI are shown in both spatial and temporal domains. The spatial intensity distribution of the total field and GPI sidebands are calculated. Numerically and experimentally obtained beam shapes of first GPI features a Gaussian-like beam profile. Our numerical results verify the unique feature of GPI and generated sidebands preserve their inherited spatial intensity profile from the input pulse for different propagation distances particularly for focused and spread the total field inside the GRIN MMF. 
\end{abstract}


%
\IEEEpeerreviewmaketitle

\section{Introduction}
Multimode optical fibers are commonly used in real world applications such as telecommunications, imaging, beam delivery due to the large core size and low nonlinearity. Multimode fibers are generally considered as an unpredictable and random light propagation environments. With recent experimental and theoretical studies, this widely acclaimed attitude is questioned. Plöschner et al. demonstrated that multimode fibers have predictable behaviors and the transferred light stays deterministic \cite{ploschner2015seeing}. Poletti et al. exploited the nonlinear interactions taking place inside of multimode fibers and study supercontinuum generation numerically \cite{poletti2008description,poletti2009dynamics,horak2012multimode}. However, with these astonishing features, multimode fibers require intricate treatments to understand complex nonlinear dynamics. 

Nowadays, GRIN MMFs, special cases for standard MMFs, are attracting great interest. With the parabolic profile of their refractive index, GRIN MMFs provide novel features. In the last few years, researchers exploited these features and reported new nonlinear dynamics to explore such as GPI \cite{krupa2016observation,wright2016self}, supercontinuum generation \cite{lopez2016visible,krupa2016spatiotemporal}, self-beam cleaning \cite{lopez2016visible,krupa2016spatial,liu2016kerr}, multimode solitons \cite{renninger2013optical} and their dispersive waves \cite{wright2015controllable}. Among these spatio-temporal effects, GPI, called also as spatio-temporal modulation instability in the literature, excels as new wavelength generation technique since generated GPI sidebands have remarkable frequency shift and inherit the spatial beam shape of pump pulse \cite{krupa2016observation,wright2016self}.

In 2003, Longhi's  theoretical work predicted GPI effect in multimode fibers \cite{longhi2003modulational}. Because of the periodic refocusing of the beam, while propagating in GRIN MMF, quasi-phase matching (QPM) (between the pump, signal and idler) resulted as GPI sidebands and discrete peaks appear in the spectrum. In contrast to intermodal four-wave mixing (FWM) which is capable of generating spectral peaks with the same amount of frequency shifts, GPI peaks inherit the spatial mode profile of the pump source \cite{nazemosadat2016phase}. Krupa et al. reported the first experimental observation of GPI sidebands by using 900 picosecond pulses with 50 kW peak power inside of GRIN MMF and observed GPI sidebands are detuned more than 120 THz from the pump frequency \cite{krupa2016observation}. This study later verified by Lopez-Galmiche et al. \cite{lopez2016visible} while demonstrating of supercontinuum generation in GRIN MMF. Their results indicated for a relatively long GRIN MMF cascaded generation of GPI sidebands evolve through the formation of a supercontinuum with contributions of higher-order nonlinear effects such as self-phase modulation (SPM), stimulated Raman scattering (SRS) and FWM generations. Very recently, Wright et al. \cite{wright2016self} studied the complex background of the self-beam cleaning and GPI sideband generation in GRIN MMF. Their detailed study contains the contributions of higher-order modes to GPI sideband generation and effect of disorder in such a nonlinear system.

Aforementioned studies about on spatio-temporal instability focused on quasi-continuous pulse (hundred ps to few ns) evolution in GRIN MMF at normal dispersion regime due to the analogy between GPI and modulation instability presented by Longhi's theoretical work \cite{longhi2003modulational}. Thus GPI dynamics for femtosecond pulses remained unknown. Here, we present first experimental investigations of the spatio-temporal instability of ultrashort pulses in GRIN MMF at normal dispersion. Femtosecond, linearly polarized pulses at 800 nm start to experience spatio-temporal evolution inside 2.6 m GRIN MMF with 50 $\mu$m core diameter and their evolution resulted as the generation of first GPI stokes and anti-Stokes pair in the experiment. Observed GPI sidebands appear in the spectrum as 91 THz detuned with respect to launched pump pulse's central frequency. Formation and the broadening tendency of GPI sidebands with increasing launch pulse energy are reported. Spatial beam shape of first GPI Stokes is measured and features Gaussian-like near-field beam profile. Theoretical calculations and numerical simulations confirm the experimental observations on the GPI sidebands. Simulation results provide detailed information on the generation and the formation behaviors of GPI sidebands. The positions of GPI sidebands in the frequency domain are within the reach of experimental observations. Furthermore, numerical studies provide information on spatial distribution of total field and GPI sidebands inside the GRIN MMF.

\section{Theoretical and Numerical Study}
Longhi's theoretical model can be simplified and used for estimating frequency offset of GPI sidebands with the pump pulse for given GRIN MMF parameters \cite{longhi2003modulational}. During sufficient amount of higher-order mode excitation in GRIN MMF, the propagating pulses experience spatial oscillations in longitudinal direction with period $\varepsilon = \pi r / \sqrt{2\Delta }$, where $\Delta$ is the relative index difference and defined as $\Delta = (n_{co}^{2}-n_{cl}^{2})/2n_{co}^{2}$, $r$ is the core radius of GRIN MMF and $n_{co} (n_{cl})$ is maximum refractive index of fiber core (clad). The required QPM condition for pump, Stokes and anti-Stokes can be expressed as $2k_{P}-k_{S}-k_{A} =-2\pi h/\varepsilon$ where $h = 1, 2, 3, ... $. Frequency separation between GPI sidebands and pump frequency ($f_{h}$) is assumed as : 

\begin{equation}
(2\pi f_{h})^{2} = \frac{2\pi h}{\varepsilon \beta _{2}} - \frac{2n_{2}\hat{I}\omega_{0}}{c\beta _{2}}
\label{eq:refname1}
\end{equation}
where $n_{2}$ is the nonlinear Kerr coefficient, $\hat{I}$ is path averaged intensity of beam, $\omega_{0}$ is the angular frequency of the pump and $\beta _{2}$ is the fiber dispersion parameter. The nonlinear part of the Eq.(\ref{eq:refname1}) has weak effect on $f_{h}$ thus one can approximate $f_{h} = \pm \sqrt{h}f_{n}$ where $2\pi f_{n} = \sqrt{2\pi/(\varepsilon \beta _{2})} $. 

\begin{figure}[b]
\centering
\includegraphics[width=\linewidth]{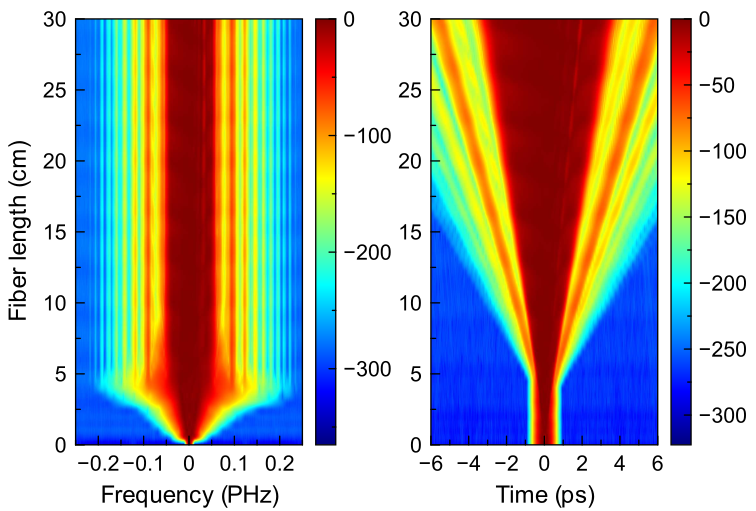}
\caption{Results from the numerical simulation with 6 cylindrically symmetric modes showing total evolution through 30 cm fiber in frequency and time domains. The intensities in dB scale.}
\label{fig:Fig1}
\end{figure}

According to this simplified model, a standard GRIN MMF with 50 $\mu$m core diameter, $0.01$ relative index difference and 36.16 ${fs^{2}}/{mm}$ group velocity dispersion at 800 nm wavelength creates spatial oscillations with $\sim$ 555 $\mu$m period for propagating light beam. For these parameters, theoretical calculations suggest that frequency offset between first GPI sideband pair and the pump pulse is about 89 THz. Regarding the pump wavelength 800 nm, expected first GPI Stokes and anti-Stokes could be located around 1049 nm and 646 nm, respectively.

\begin{figure}[b]
\centering
\includegraphics[width=\linewidth]{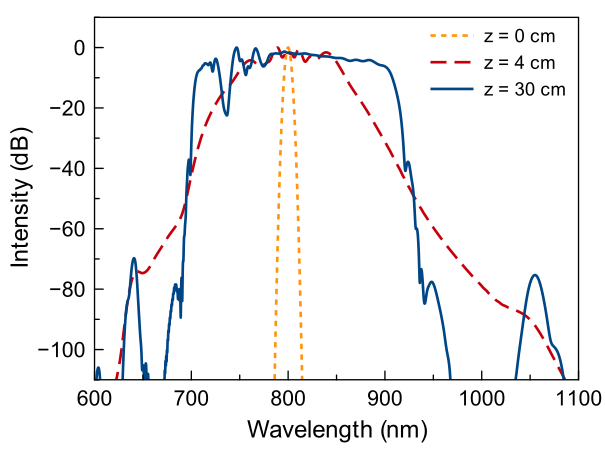}
\caption{Numerical results for spectral evolution inside the GRIN MMF with 50$\mu$m core diameter. Beam profiles are in $\mu$m scale.}
\label{fig:Fig2}
\end{figure}

\begin{figure*}[t!]
\centering
\includegraphics[width=1.0\textwidth]{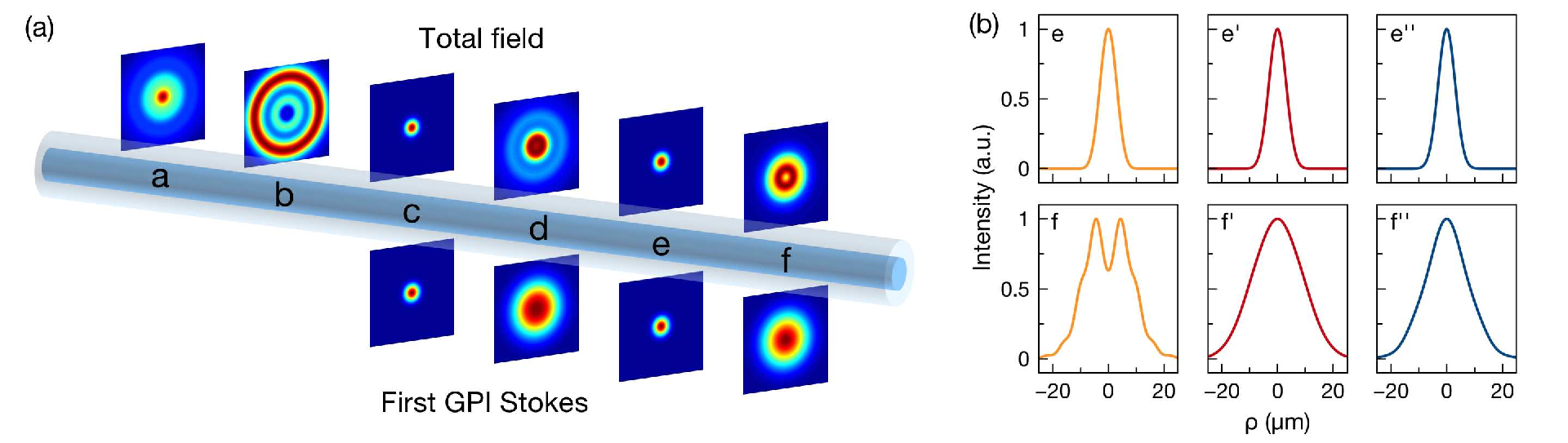}
\caption{Numerical results for spatial evolution inside the GRIN MMF with 50$\mu$m core diameter. (a) Spatial intensity distributions at 2 cm, 5 cm, 17.03 cm, 17.12 cm, 17.21 cm, 17.28 cm (a-f). (b) Beam profiles of total field (e, f), first GPI Stokes (e', f') and first GPI anti-Stokes (e'', f'') for 17.21 and 17.28 cm of the GRIN MMF, respectively.}
\label{fig:Fig3}
\end{figure*}

Theoretical calculations are derived independently from the pump pulse duration. To study the detailed evolution of femtosecond pulses in GRIN MMF numerical simulations are needed. Pulse propagation in GRIN MMF can be simulated using the generalize multimode nonlinear Sch\"{o}dinger equation \cite{poletti2008description,horak2012multimode,mafi2012pulse}.
According to this model, the complex electric field can be expanded into a sum for modes p = 0,1,2,\dots and each mode can be represented with a transverse fiber mode profile. Evolution of temporal envelope of pth mode can be written as:
\begin{equation}
\begin{split}
\frac{\partial A_{p}}{\partial z} = i\delta \beta _{0}^{(p)}A_{p}-\delta \beta _{1}^{(p)}\frac{\partial A_{p}}{\partial t}
-i\frac{\beta_{2}^{(p)} }{2}\frac{\partial^2 A_{p}}{\partial^2 t} \\
+i\frac{\gamma}{3}(1+\frac{i}{\omega_{0}}\frac{\partial}{\partial t})
\sum_{l,m,n}\eta_{plmn}[(1-f_{R})A_{l}A_{m}A_{n}^{*} \\
+f_{R}A_{l}\int h_{R}A_{m}(z,t-\tau)A_{n}^{*}(z,t-\tau)d\tau]
\label{eq:refname2}
\end{split}
\end{equation}
where $\eta_{plmn}$ is nonlinear coupling coefficient, $f_{R} \approx 0.18$ is the fractional contribution of the Raman effect,$h_{R}$ is the delayed Raman response function and $\delta \beta _{0}^{(p)}$ ($\delta \beta _{1}^{(p)}$) is difference between first (second) Taylor expansion coefficient of propagation constant for corresponding and the fundamental mode. Propagation constants for each mode ($p$) can be written as:

\begin{equation}
\beta_{p}(\omega)=\frac{\omega n_{0}(\omega)}{c}\sqrt{1-\frac{2c}{\omega n_{0}(\omega)}\frac{\sqrt{2\Delta }}{r}(2p+1)}
\label{eq:refname3}
\end{equation}
where $n_{0}$ is the refractive index of fiber core, $\omega$ is frequency and $c$ is the speed of light in vacuum. Dispersion curve and group velocity dispersion (GVD) for each mode can be derived from Eq.(\ref{eq:refname3}). To solve Eq.(\ref{eq:refname2}) numerically, we use symmetrized split-step Fourier method \cite{agrawal2007nonlinear} and include Raman process and shock terms in our simulations. 

A GRIN MMF with 50 $\mu$m core diameter supports approximately 415 modes at 800 nm and simulating all of them will require complex and time-consuming calculations. Thus to achieve manageable computation times, we consider only first six zero-angular-momentum modes in our simulation. We launch pulses with 200 fs pulse duration, 350 nJ pulse energy at 800 nm which has a peak power considerably below the critical power for Kerr-induced self-focusing ($\sim$ 2.44 MW) and this initials pulse energy is distributed among these six modes (50\% in p=0, 18\% in p=1, 13\% in p=2, 10\% in p=3, 6\% in p=4 and 3\% in p=5). Propagation constants, dispersion and nonlinearity parameters are calculated as in \cite{poletti2008description,horak2012multimode,mafi2012pulse}. We set $n_{0}$ as 1.4676, $n_{2}$ as $2.7x10^{-20} m^{2}/W$, relative index difference as $0.01$, integration step as 10 $\mu$m, time window width as 15 ps with 2 fs resolution.

Spectral and temporal evolution of the pump pulse is first studied numerically for 30 cm GRIN MMF  (Fig. \ref{fig:Fig1}). Femtosecond pulse starts to broaden in frequency domain while propagating in the very first part of the GRIN MMF. The observed relatively large spectral broadening is a unique feature of GRIN MMF and caused by high pump pulse energy \cite{manassah1988self,karlsson1992dynamics}. Numerical results indicate the generation of GPI sidebands requires approximately 100 oscillations inside the GRIN MMF for our launch conditions and parameters (pulse duration, peak power and fiber core size etc.). Broadened pulse covers the emergence of GPI sidebands and after a certain amount of propagation, discrete peaks become obviously visible. In this aspect, generation of GPI sidebands from ultrashort pump pulses and SPM induced modulation-instability possess similar characteristics in frequency domain \cite{agrawal2007nonlinear}. Formation of GPI peaks in femtosecond regime is different than the formation in nanosecond regime, in which the GPI sidebands emerge directly from noise level without assisted SPM broadened pump pulses \cite{krupa2016observation}. In numerical results, first pairs of GPI sidebands appear at frequencies detuned approximately 90 THz from the launched pump pulse frequency. As shown in Fig.\ref{fig:Fig2}, first GPI Stokes and anti-Stokes are centered around 1055 nm and 640 nm, respectively. Simulation results indicate that after the formation is established, the intensity of GPI sidebands starts to increase due to constant frequency generation with spatio-temporal propagation. Along the fiber, the positions of GPI sidebands at frequency domains remains stable.

Obtaining the spatial evolution of the pulse inside the fiber is important in order to understand spatio-temporal changes. Thus we calculate the spatial evolution numerically at various positions inside the fiber and presented in Fig. \ref{fig:Fig3} according to the simulation model \cite{horak2012multimode,renninger2013optical,mafi2012pulse}. Our results verify the spatial evolution of the beam experiences periodic refocusing along the GRIN MMF. This periodic behavior preserves the Gaussian-like spatial distribution for all focused points. We show the beam profile for three different focused points (Fig. \ref{fig:Fig3}(a).a, Fig. \ref{fig:Fig3}(a).c and Fig. \ref{fig:Fig3}(a).e). During the GPI sideband generation ($\sim$ 5 cm), we observed non-Gaussian intensity distribution for total field Fig. \ref{fig:Fig3}(a).b. After the GPI sideband generation occurs, spatial intensity distribution of total field approaches to Gaussian beam shape for spread points Fig. \ref{fig:Fig3}(a).d and Fig. \ref{fig:Fig3}(a).f as well. From numerical calculations, we also extract spatial intensity distribution of first GPI Stokes and it indicates that inherited spatial intensity distribution is preserved at focused and spread points Fig. \ref{fig:Fig3}(a).c-f. This observation is a unique feature of the spatio-temporal evolution of the GPI in GRIN MMF. In addition, we also studied spatial evolution of first GPI anti-Stokes in different positions (focused and spread) along the GRIN MMF. As presented in Fig. \ref{fig:Fig3}(b), Gaussian-like spatial distributions are preserved for first GPI anti-Stokes as well.

\begin{figure}[t!]
\centering
\includegraphics[width=\linewidth]{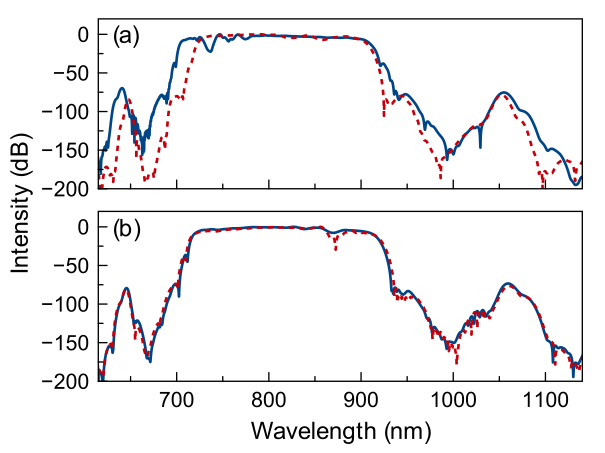}
\caption{Numerical spectra obtained from 30 cm GRIN MMF with different parameters. (a) Calculations with 6 cylindrically symmetric modes for different initial energy distributions (b) calculations with 3 cylindrically symmetric modes for different initial energy distributions.}
\label{fig:Fig4}
\end{figure}

We study the effect of different launch conditions on GPI sideband generation in the Fig.\ref{fig:Fig4}(a). We compare above mentioned result (solid line) with the initial energy distribution between the modes as 30\% in p=0, 25\% in p=1, 15\% in p=2, 5\% in p=3, 3\% in p=4 and 2\% in p=5 (dashed line). Decreasing the energy of fundamental mode (p=0) results in less spectral broadening and slight frequency shift to first anti-Stokes sideband. Intensity difference between first GPI sidebands between the different launch energy distribution is observed as 4 dB. Next, we check the impact of considered number of modes in numerical studies. For longer fiber lengths one may need to decrease number of simulated modes to further reduce calculation times. For this reason, we run simulations with only first three zero-angular-momentum modes with different initial energy distributions. In the literature, it is shown that simulations with three lowest order modes reflect sufficiently similar results with experiments \cite{renninger2013optical} thus this approach is an acceptable simplification. Obtained spectra for 30 cm GRIN MMF are presented in the Fig.\ref{fig:Fig4}(b). First we distribute launch energy as 50\% in p=0, 30\% in p=1, 20\% in p=2 (solid line) then change as 35\% in p=0, 35\% in p=1, 30\% in p=2 (dashed line). Obtained spectra for these energy distributions present nearly identical features.

\section{Experimental Results and Discussions}

In the experiments, we use amplified Ti:Sapphire laser (Spitfire by Spectra-Physics) capable to generate linearly polarized, single-mode, 200 femtosecond ultrashort pulses at 800 nm with 1 kHz repetition rate for pump source. The fiber used in the experiment is a commercially available GRIN MMF (Thorlabs-GIF50C) with 50 $\mu$m (125 $\mu$m) core (clad) diameter and 0.2 numerical aperture. We couple pump pulses into 2.6 m GRIN MMF with plano-convex lens and three-axis translation stage configuration. We test various lenses with different focal lengths and obtain GPI sideband generation in the measured spectrum with beam waists on fiber facet. For small beam waists, observed GPI sidebands are unstable due to an environmental issue such as vibrations and degradation of optical alignment tools. On the other hand, selected 60 mm focal length lens provides $\sim$ 20 $\mu$m waist size, thus excitation of higher-order modes is obtained easily and measured GPI sidebands remain stable for several hours. In general, free space coupling efficiency greater than 80\% could be achieved. 

\begin{figure}[b!]
\centering
\includegraphics[width=\linewidth]{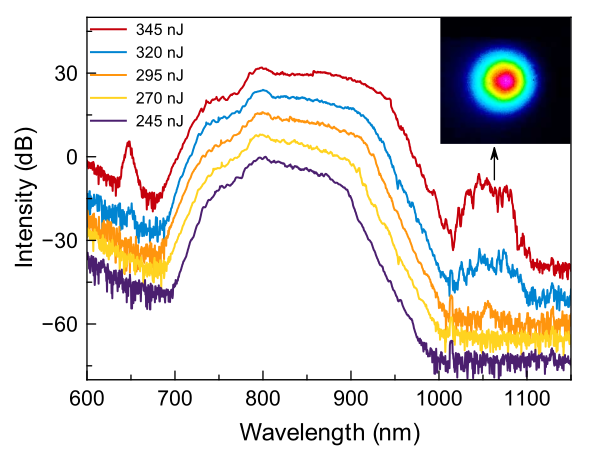}
\caption{Measured spectra as a function of launched pulse energy. Inset: near-filed beam profile of 1. GPI Stokes.}
\label{fig:Fig5}
\end{figure}

The generation and formation of experimentally observed GPI sidebands are reported in detail Fig. \ref{fig:Fig5} for different launched pulse energy conditions. We launch 200 fs pump pulses at 800 nm with $\sim$ 10 nm FWHM into a 2.6 m GRIN MMF. Pump pulse experiences asymmetric spectral broadening for relatively low pulse energies (for example 270 nJ). The observed asymmetric spectral broadening could be the result of stimulated Raman scattering (SRS). At high launched pulse energy (345 nJ), we observe further spectral broadening on pump region but discrete SRS peaks formation is not detected. With the increasing launched pulse energy for constant fiber length, first GPI Stokes sideband emerges at 295 nJ launched pulse energy. Theory and simulation results indicate that both GPI sidebands should appear at the same time. Thus for launched pulse energy of 295 nJ, first GPI anti-Stokes should lie under the noise level of the optical spectrum analyzer. As launched pulse energy increases (at 320 nJ), amplification and spectral broadening for first GPI Stokes is recorded and in addition, first GPI anti-Stokes also emerges. Similar spectral evolution (amplification and broadening) is also obtained for first anti-Stokes.

We successfully generate first GPI peak pair (first Stokes and anti-Stokes) with launched 345 nJ femtosecond pulses into GRIN MMF. First GPI peak pair is observed with $\sim$ 91 THz separation with respect to pump frequency $(f_{0})$ (see Fig. \ref{fig:Fig4}(a)). First Stokes and anti-Stokes are centered around 1055 nm and 645 nm, respectively. The corresponding optical spectrum bandwidth of first Stokes and anti-Stokes are $\sim$ 12 nm and $\sim$ 5 nm, respectively. Even though bandwidths of GPI sidebands seem different in wavelengths, in frequency domain they have approximately close bandwidths of 3.2 THz. We measure the near-field beam profile of the first GPI Stokes sideband for launched pump pulses of 345 nJ pulse energy ( Fig.\ref{fig:Fig5}-inset). To separate first GPI Stokes from the pump pulse and the GPI anti-Stokes, we use a longpass filter with 1000 nm cutoff wavelength. As expected from a GPI sideband, a clean (speckle free), Gaussian-like near-field beam profile is observed which is similar to the pump beam shape. This feature of GPI sidebands is the signature of GPI in GRIN MMF and significant difference between GPI and intermodal FWM. \cite{krupa2016observation,lopez2016visible}. 

\begin{figure}[t!]
\centering
\includegraphics[width=\linewidth]{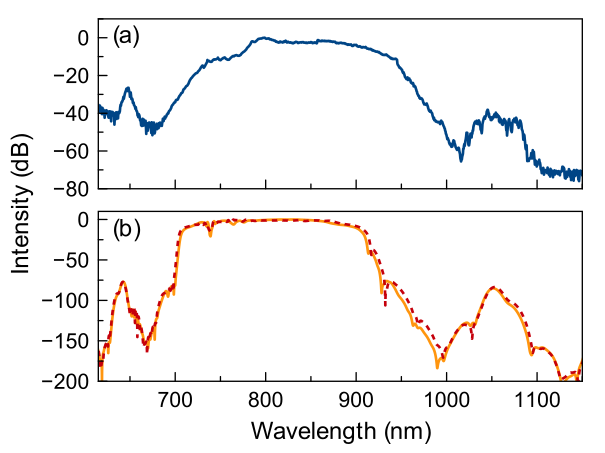}
\caption{Optical spectra obtained after propagating 2.6 m GRIN MMF. (a) Experimental measurement and (b) simulation results with 3 cylindrically symmetric modes for different energy distribution between the modes.}
\label{fig:Fig6}
\end{figure}

To compare obtained experimental result with our numerical model, we perform numerical simulations for 2.6 m GRIN MMF (see Fig. \ref{fig:Fig6}). In order to decrease simulation time to manageable durations we simulate first three zero-angular-momentum modes with included Raman process and shock terms. First, we distribute 345 nJ pulse energy of the launched pulse to modes such as 50\% in p=0, 30\% in p=1 and 20\% in p=2 (solid-line). To check the effect of energy distribution between the modes we run simulations with 35\% in p=0, 35\% in p=1 and 30\% in p=2 distribution as well (dashed-line). For both distributions, positions and bandwidths of GPI sidebands on frequency domain are similar to experimentally obtained results. The low-intensity level of GPI sidebands in simulations may arise from neglecting the contribution of higher-order modes. 

\section{Conclusion}
In conclusion, we study the spatio-temporal instability of ultrashort pulses in normal dispersion GRIN MMF. Our experimental results present GPI formation with femtosecond pump pulses first time in the literature and the reported GPI sidebands are well-aligned with theoretical predictions and numerical calculations. Detailed numerical studies revealed the generation and propagation behaviors of GPI sidebands inside of GRIN MMF. Observed results provide inside of the GPI formation with ultrashort pulses to complete spatio-temporal pulse evolution studies and indicate that the known attractor effect observed in GRIN MMF for quasi-continuous pump pulses also exists for ultrashort pump pulses \cite{wright2016self}. This attractor causes first self-beam cleaning then propagating beam experience GPI which manifest itself with sidebands in the frequency domain. For femtosecond pulses, self-beam cleaning is presented by Liu et al. \cite{liu2016kerr} and our result complete the information gap for recently emerging research field. We report that, from certain aspects, the generation mechanism of GPI sidebands is analogous to SPM assisted FWM in standard single mode fiber. Experimental observations show that first GPI Stokes features Gaussian-like spatial intensity profile. Numerically calculated spatial intensity profiles verify experimental measurements and provides information on spatial evolution of the beam inside the GRIN MMF. With intrinsic large frequency shift, GPI sidebands can be employed to generate new wavelengths for various application purposes. In future direction, the presented spatio-temporal platform provides potential directions to investigate its complex, nonlinear dynamics with ultrashort pulses such as self-beam cleaning and supercontinuum generation.


%

\section*{Acknowledgment}
The authors thank \c{C}. \c{S}enel for discussions and The Turkish Academy of Sciences — Outstanding Young Scientists Award Program (TUBA-GEBIP); Bilim Akademisi — The Science Academy, Turkey under the BAGEP program; METU Prof. Dr. Mustafa Parlar Foundation; FABED for their supports.

\ifCLASSOPTIONcaptionsoff
  \newpage
\fi


\bibliographystyle{IEEEtran}  
\bibliography{main}

%

\begin{IEEEbiography}[{\includegraphics[width=1in,height=1.25in,clip,keepaspectratio]{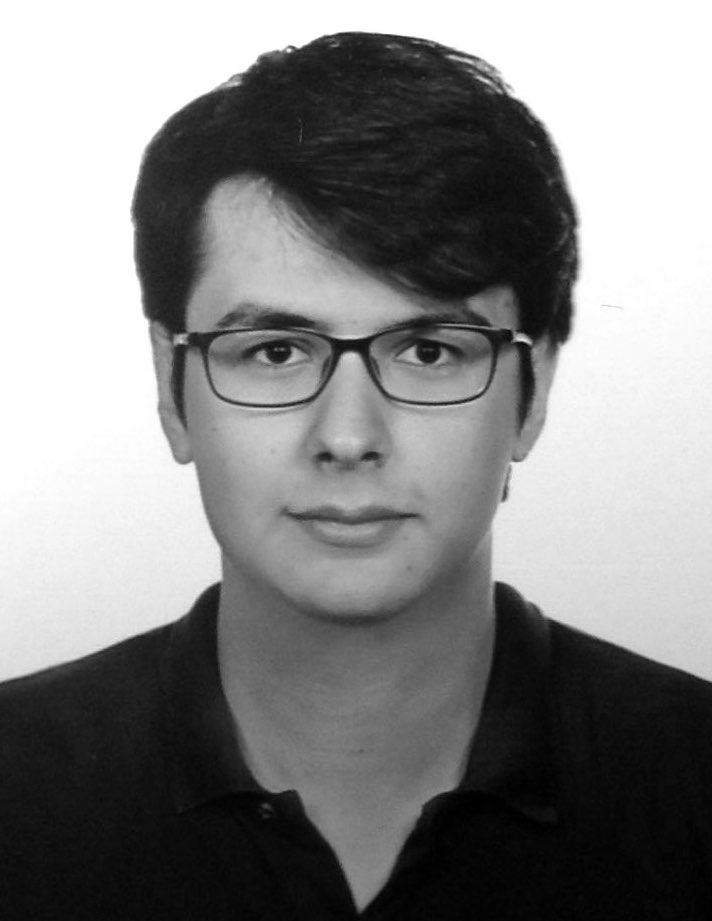}}]{U\u{g}ur Te\u{g}in}
 was born in Bursa, Turkey in 1992. He received the B.S. degree in physics from the Bilkent University, Turkey in 2015. His research interests include mode-locked lasers, fiber amplifiers, nonlinear fiber optics and  spatio-temporal nonlinear dynamics. He is currently pursuing the master’s degree in Materials Science and Nanotechnology at Bilkent University, Turkey. He is a member of the Optical Society of America and SPIE. 
\end{IEEEbiography}

\begin{IEEEbiography}
[{\includegraphics[width=1in,height=1.25in,clip,keepaspectratio]{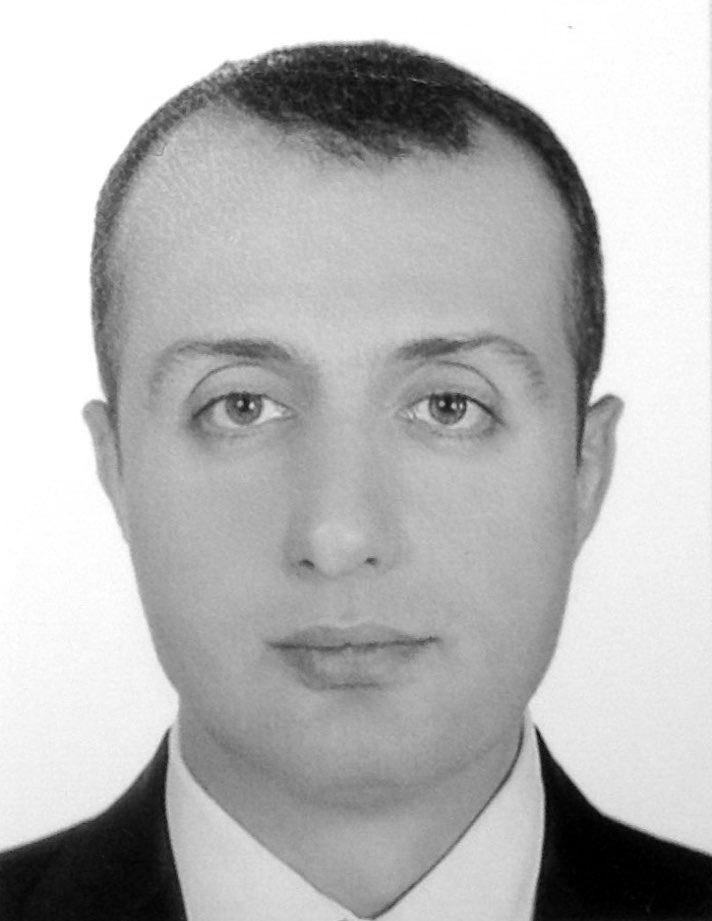}}]{B\"{u}lend Orta\c{c}}
 received the B.S. degree in Physics from the Karadeniz Technical University, Trabzon, Turkey, in 1997, M.S. degree in Teaching and Diffusion of Sciences and Technology from ENS Cachan University, Paris, France, in 2000, and Ph. D. degree in Optoelectronics from Rouen University, Rouen, France, in 2004 respectively. In Mach 2005, he joined the Institute of Applied Physics, Friedrich-Schiller University, Jena, Germany, as a Post-Doctoral Associate. Since November 2009, he has been working as a research assistant professor at Institute of Materials Science and Nanotechnology, Bilkent University. He is the founder and principle investigator of the Laser Research Laboratory. His current research interests include the development of powerful fiber lasers in the continuous-wave regime to pulsed regime (ns, ps and fs) and the demonstration of laser systems for real world applications. He has published more than 100 research articles in major peer-reviewed scientific journals (over 60) and conferences (over 150) in the field of laser physics.
\end{IEEEbiography}


\vfill


\end{document}